\documentclass[12pt]{article}
\usepackage{amsfonts}
\usepackage{amsmath}
\usepackage{amssymb}
\usepackage{axodraw}
\usepackage{mathptmx}
\usepackage[dvips]{color}
\usepackage{epsfig}
\usepackage{graphicx}
\newcommand{\calC}{{\cal C}}
\newcommand{\calL}{{\cal L}}

\newcommand{\Br}{\textrm{Br}}
\newcommand{\eV}{{\rm eV}}
\newcommand{\GeV}{{\rm GeV}}
\newcommand{\cm}{{\rm cm}}

\begin{document}
\baselineskip=16pt

\pagenumbering{arabic}

\vspace{1.0cm}

\begin{center}
{\Large\sf Electric dipole moments of charged leptons from their
Majorana-type Yukawa couplings}
\\[10pt]
\vspace{.5 cm}

{Yi Liao\footnote{liaoy@nankai.edu.cn}}

{Department of Physics, Nankai University, Tianjin 300071, China}

\vspace{2.0ex}

{\bf Abstract}

\end{center}

The electric dipole moments (EDMs) of charged leptons are
significantly suppressed in standard model. It has been found
previously that they are even more severely suppressed in seesaw
type models by powers of tiny neutrino masses as far as a leptonic
CP source is concerned. We investigate whether a Majorana-type
Yukawa coupling between charged leptons and a doubly charged
scalar can contribute significantly to their EDMs. An observable
EDM would then help unravel the Majorana nature of neutrinos by a
lepton number conserving quantity. We find that the EDMs are
indeed parametrically large, of the form $d_\alpha\propto
em_\alpha(m^2_\beta-m^2_\gamma)/m^4$ up to logarithms, where
$m_\alpha$ and $m$ are respectively the masses of charged leptons
and the scalar. And they satisfy a sum rule to good precision,
$d_e/m_e+d_\mu/m_\mu+d_\tau/m_\tau=0$. With the most stringent
constraints from lepton flavor violating transitions taken into
account, their values are still much larger than the mentioned
previous results. Unfortunately, even in the most optimistic case
the electron EDM is about three orders of magnitude below the
foreseeable experimental sensitivity.

\begin{flushleft}
PACS: 13.40.Em, 14.60.-z, 12.60.Fr, 14.60.Pq

Keywords: electric dipole moment, Majorana neutrino, doubly charged
scalar

\end{flushleft}

\newpage

\section{Introduction}
\label{sec:intro}

The CP violation through a Dirac phase in the CKM matrix of weak
interactions has been well tested in the flavored systems of
hadrons. It is generally believed however that this cannot be the
unique or even the dominant source of CP violation because of the
observed large baryon number asymmetry in our universe (BAU). One
of attractive solutions to BAU is offered by the mechanism of
leptogenesis \cite{Fukugita:1986hr} in which the lepton number
asymmetry is first generated through CP violation in the lepton
sector and then converts partly into BAU via sphaleron effects
\cite{Kuzmin:1985mm}.

Our current information on the leptonic mixing matrix comes
dominantly from experiments of neutrino oscillations
\cite{Maltoni:2007zf}. The matrix is CKM-like involving a single
Dirac phase if neutrinos are Dirac particles, but can contain
additional two CP phases if neutrinos are of Majorana nature. The
oscillations are blind to the latter Majorana phases while their
sensitivity to the Dirac phase is seriously diminished by a very
small, if not vanishing, mixing angle out of three. This leaves CP
violation in the lepton sector largely untested so far except
perhaps for the experiment of neutrinoless double beta ($0\nu
2\beta$) decays which can be sensitive to CP phases but whose
status is under debate.

Nevertheless, there is another physical observable, the electric
dipole moment (EDM), that can provide an independent probe to CP
violation. The current experimental limits on the EDMs of the
mercury atom \cite{Romalis:2000mg}, neutron \cite{Baker:2006ts},
electron \cite{Regan:2002ta} and muon \cite{Bennett:2008dy} are
already very impressive, and further improvements are expected to
take place in the near future \cite{Pospelov:2005pr}. These EDMs can
in principle be induced by the Dirac phase in the CKM matrix of
standard model (SM). However, it was known long ago that the
electric and chromoelectric dipole moments of quarks vanish to the
two-loop order \cite{Shabalin:1978rs}. This was interpreted as a
joint result of two features in SM \cite{Liao:1999yt}, namely the
unitarity of the CKM matrix and the purely left-handed chirality of
the charged current, and relaxation of any of them would yield quark
EDMs at two loops \cite{Liao:1999fc}. The lepton EDMs then become
extremely small in SM as they are first induced at four loops
\cite{Pospelov:1991zt}. This makes them a potentially ideal place to
search for CP violation in the lepton sector.

If neutrinos are Dirac particles, the EDMs of charged leptons will
be hopelessly tiny. The case of quarks in SM repeats in the lepton
sector in an even worse manner since neutrinos can be considered
degenerate to very good precision at the weak scale, in which case
there is effectively no CP violation in the lepton sector. But the
situation could be different when neutrinos are Majorana particles
because of peculiarities with Majorana CP phases
\cite{deGouvea:2002gf}. Since Majorana phases dominate over the
Dirac phase in this case, an observable lepton's EDM would not only
discover CP violation in the lepton sector but also expose the
Majorana nature of neutrinos by a lepton number conserving quantity
in sharp contrast to $0\nu 2\beta$ decays. Indeed, as pointed out in
\cite{Ng:1995cs}, there is a topologically new type of two-loop
Feynman diagrams when neutrinos are Majorana particles that can
contribute to the charged lepton EDMs. But it was found subsequently
that this type of contribution is always severely suppressed by
neutrino masses from virtual loops whether one works in the standard
type I seesaw model \cite{Archambault:2004td} or one augmented with
an additional Higgs doublet \cite{Chang:2004pba}, or in type II
seesaw \cite{deGouvea:2005jj}. The obtained numbers are actually
even smaller than the four-loop result due to the CKM phase, and
thus would not be observable in any foreseeable experiments.

When neutrinos are Majorana particles, the lepton number may be
violated either by a bare Majorana-type mass of heavy neutrinos that
are singlets of SM, or by some other fields that are active in SM
and couple in particular to leptons. The physics at low energies is
much richer in the latter case. And the simplest choice would be to
add a scalar triplet as in the type II seesaw model
\cite{Konetschny:1977bn}. We are thus motivated to consider the most
general Majorana-type Yukawa couplings of charged leptons to a
doubly charged scalar. These couplings could arise as, but are not
restricted to, part of interactions as in type II seesaw or a larger
extension of SM. The CP violation encoded in the couplings can
induce EDMs to charged leptons, and we find that this contribution
is indeed parametrically large. Besides the product of Yukawa
couplings and a lepton mass factor for chirality flip, the EDM is
suppressed by charged lepton masses squared over four powers of the
scalar mass and is partly enhanced by a logarithm. In particular, it
incurs no suppression by neutrino masses since no neutrinos appear
in virtual loops. This is the largest term in lepton EDMs, to our
knowledge, coming from a flavor dependent CP source.

The paper is organized as follows. In the next section we describe
the Majorana-type Yukawa couplings between the charged leptons and
doubly charged scalars, and count the number of independent
physical parameters. The two-loop diagrams for EDMs are then
evaluated analytically in section 3, and a sum rule is found.
Using the most stringent constraints from lepton flavor violating
decays we estimate in section 4 the largest allowed values for the
EDMs. We summarize and conclude in the last section.

\section{Majorana-type Yukawa couplings}

The relevant interactions for our study are the Majorana-type Yukawa
couplings
\begin{eqnarray}
\calL_{\rm Yuk}=\ell^Tb\calC P_L\ell\xi^{++}+\bar\ell b^\dagger\calC
P_R\bar\ell^T\xi^{--} %
\label{eqn_yukawa}
\end{eqnarray}
and the standard QED
\begin{eqnarray}
\calL_{\rm QED}=-eA^\mu\bar\ell\gamma_\mu\ell%
+2ieA^\mu(\xi^{--}\partial_\mu\xi^{++} -
\xi^{++}\partial_\mu\xi^{--})%
\label{eqn_qed}
\end{eqnarray}
Here $\ell$, $\xi^{\pm\pm}$ and $A_\mu$ are respectively the charged
lepton, doubly charged scalar and electromagnetic fields, and $e$ is
the electromagnetic coupling. We use Greek letters to denote the
three charged leptons. $\calC=i\gamma^0\gamma^2$ is the matrix
employed in charge conjugation and
$P_{R,L}=\frac{1}{2}(1\pm\gamma_5)$ are chiral projectors.

The Yukawa coupling matrix $b$ is symmetric in lepton flavors due to
antisymmetry in fermion fields but is otherwise arbitrary. With $n$
flavors of leptons, $b$ has generally $n+\frac{1}{2}n(n-1)$ moduli
and $n+\frac{1}{2}n(n-1)$ phases. All moduli are physical
parameters. However, not all of the phases are physical. For
instance, the phases in the diagonal entries $b_{\alpha\alpha}$ can
all be removed by redefining complex fields $\ell_\alpha$. After
this, there are no more degrees of freedom to rephase fields without
reintroducing phases into $b_{\alpha\alpha}$. There are thus only
$\frac{1}{2}n(n-1)$ physical phases. They signal $T$ and $CP$
violation as we analyze below.

When the matrix $b$ is real, we can prescribe the $T$ and $CP$
transformations as $T\xi^{++}T^{-1}=+\xi^{++}$,
$(CP)\xi^{++}(CP)^{-1}=-\xi^{--}$ so that both are preserved by the
above interactions. If $b$ is purely imaginary, we can prescribe in
the opposite manner to preserve both $T$ and $CP$,
$T\xi^{++}T^{-1}=-\xi^{++}$, $(CP)\xi^{++}(CP)^{-1}=+\xi^{--}$. The
latter case can of course be reduced to the former by rephasing the
$\xi^{++}$ field by a factor of $i$. Therefore, $T$ and $CP$
symmetries are violated only when the matrix $b$ is genuinely
complex, neither real nor purely imaginary.

The above results are general and do not rely on any model. It is
also possible to preserve lepton number by assigning two units to
$\xi^{--}$. Our later calculation of EDM applies to the general
case. But since the doubly charged scalars appear naturally in type
II seesaw, it is interesting to consider this particular case
separately:
\begin{eqnarray}
b=\frac{1}{2v_3}V^\ast m_\nu V^\dagger %
\label{eqn_seesaw}
\end{eqnarray}
where $V$ is the lepton mixing matrix and $m_\nu$ the diagonal
neutrino mass matrix with real, semi-positive eigenvalues $m_i$.
The vacuum expectation value of the scalar triplet, $v_3$, is
induced from that of the scalar doublet through a soft lepton
number violating term. It is possible and common practice in type
II seesaw to arrange order one Yukawa couplings $b$ by assuming
$v_3$ to be the same order of magnitude as $m_\nu$. The moduli in
$b$ correspond to $n$ neutrino masses (over $|v_3|$) plus
$\frac{1}{2}n(n-1)$ mixing angles in $V$, while the physical
phases are equivalent to $\frac{1}{2}(n-1)(n-2)$ Dirac phases and
$(n-1)$ Majorana phases in $V$.

\section{Evaluation of electric dipole moments}

Now we calculate the EDM $d_\alpha$ induced for the charged lepton
$\ell_\alpha$ due to interactions in eqs. (\ref{eqn_yukawa},
\ref{eqn_qed}). The effective EDM interaction is defined as
\begin{eqnarray}
\calL_{\rm EDM}=-\frac{i}{2}d_\alpha
\bar\ell_\alpha\gamma_5\sigma_{\mu\nu}\ell_\alpha F^{\mu\nu}%
\label{eqn_edm}
\end{eqnarray}
There is no contribution at one loop level since the matrix
element $b_{\alpha\beta}$ always appears in a self-conjugate form,
$|b_{\alpha\beta}|^2$, so that no phases can survive. The other
way to see this is to notice that, when computing $d_\alpha$ for a
specific $\alpha$, one can choose suitable phases for the
$\ell_\beta$ fields so that all of $b_{\alpha\beta}$ are real.
Thus more factors of $b$ have to be involved to induce an EDM, and
the first contribution occurs at two loop level.


The two-loop Feynman diagrams contributing to $d_\alpha$ are
depicted in Fig. 1. The incoming and outgoing momenta of the
$\ell_\alpha$ are respectively $p\pm\frac{1}{2}q$ with $q$ being
the outgoing momentum of the photon attached at the vertex
indicated by $\otimes$. The arrows in the graphs denote the flow
of negative charges, and the summation over the virtual charged
leptons $\ell_\beta$, $\ell_\gamma$ and $\ell_\delta$ is implied.
We find that because of the chirality structure in $\calL_{\rm
Yuk}$ the chirality flip required by the EDM has to be done by the
external lepton mass, $m_\alpha$. Upon extracting out this mass
factor we ignore further dependence on it. This is a good
approximation for practical purposes with incurred relative errors
of order $O(r_\alpha)$, where $r_\alpha=m^2_\alpha/m^2$ and $m$ is
the mass of $\xi^{\pm\pm}$. The dependence on other charged lepton
masses enters in a quadratic form, i.e., via
$r_{\beta,\gamma,\delta}$.

The lepton flavor dependence in the relevant term of each graph can
thus be described as
\begin{eqnarray}
b^\ast_{\gamma\alpha}b^\ast_{\delta\beta}b_{\delta\gamma}b_{\beta\alpha}
f(r_\beta,r_\gamma;r_\delta),
\end{eqnarray}
where $f$ is a function of the indicated mass ratios. $f$ is
generally a sum of terms that are respectively symmetric and
antisymmetric in $\beta$ and $\gamma$. The symmetric term cannot
contribute to EDM since we are effectively summing the
self-conjugated $b$ factors,
$b^\ast_{\gamma\alpha}b^\ast_{\delta\beta}
b_{\delta\gamma}b_{\beta\alpha}+{\rm c.c.}$, which do not vanish
even for a real or purely imaginary $b$. The antisymmetric
combination on the other hand survives only when $b$ is genuinely
complex with CP phases involved:
\begin{eqnarray}
\frac{i}{2}\Im[b^\ast_{\gamma\alpha}b^\ast_{\delta\beta}
b_{\delta\gamma}b_{\beta\alpha}]
[f(r_\beta,r_\gamma;r_\delta)-f(r_\gamma,r_\beta;r_\delta)]
\end{eqnarray}
Since $r_\delta\ll 1$, the leading term, if not vanishing, is
obtained by setting $r_\delta=0$. The $b$ factors then degenerate
into the form
\begin{eqnarray}
\Im[b^\ast_{\gamma\alpha}b_{\beta\alpha}(b^\dagger b)_{\beta\gamma}]
\end{eqnarray}

\begin{center}
\begin{picture}(320,280)(0,0)

\SetOffset(20,220) %
\ArrowLine(-20,0)(0,0)\ArrowLine(20,0)(0,0)%
\ArrowLine(20,0)(80,0)\ArrowLine(100,0)(80,0)%
\ArrowLine(100,0)(120,0)%
\DashArrowArc(50,0)(30,0,180){3}%
\DashArrowArcn(50,0)(50,180,95){3}\DashArrowArcn(50,0)(50,85,0){3}%
\Text(-10,8)[]{$\alpha$}\Text(10,8)[]{$\beta$}
\Text(50,8)[]{$\delta$}\Text(90,8)[]{$\gamma$}
\Text(110,8)[]{$\alpha$} %
\Text(50,50)[]{$\otimes$}%
\Text(50,-20)[]{$(a)$}

\SetOffset(180,220) %
\ArrowLine(-20,0)(0,0)\ArrowLine(20,0)(0,0)
\ArrowLine(20,0)(80,0)\ArrowLine(100,0)(80,0)\ArrowLine(100,0)(120,0)%
\DashArrowArc(50,0)(30,0,80){3}\DashArrowArc(50,0)(30,100,180){3}%
\DashArrowArcn(50,0)(50,180,0){3}%
\Text(50,30)[]{$\otimes$}%
\Text(50,-20)[]{$(b)$}

\SetOffset(20,130) %
\ArrowLine(-20,0)(0,0)\ArrowLine(20,0)(0,0)
\ArrowLine(20,0)(46,0)\ArrowLine(54,0)(80,0)
\ArrowLine(100,0)(80,0)\ArrowLine(100,0)(120,0)%
\DashArrowArc(50,0)(30,0,180){3}\DashArrowArcn(50,0)(50,180,0){3}%
\Text(50,0)[]{$\otimes$}%
\Text(50,-20)[]{$(c)$}

\SetOffset(180,130) %
\ArrowLine(-20,0)(0,0)\ArrowLine(7,0)(0,0)\ArrowLine(20,0)(13,0)
\ArrowLine(20,0)(80,0)\ArrowLine(100,0)(80,0)\ArrowLine(100,0)(120,0)%
\DashArrowArc(50,0)(30,0,180){3}\DashArrowArcn(50,0)(50,180,0){3}%
\Text(10,0)[]{$\otimes$}%
\Text(50,-20)[]{$(d)$}

\SetOffset(20,40) %
\ArrowLine(-20,0)(0,0)\ArrowLine(20,0)(0,0)
\ArrowLine(20,0)(80,0)\ArrowLine(100,0)(94,0)\ArrowLine(86,0)(80,0)
\ArrowLine(100,0)(120,0)%
\DashCArc(50,0)(30,0,180){3}\DashCArc(50,0)(50,0,180){3}%
\DashArrowArc(50,0)(30,0,180){3}\DashArrowArcn(50,0)(50,180,0){3}%
\Text(90,0)[]{$\otimes$}%
\Text(50,-20)[]{$(e)$}

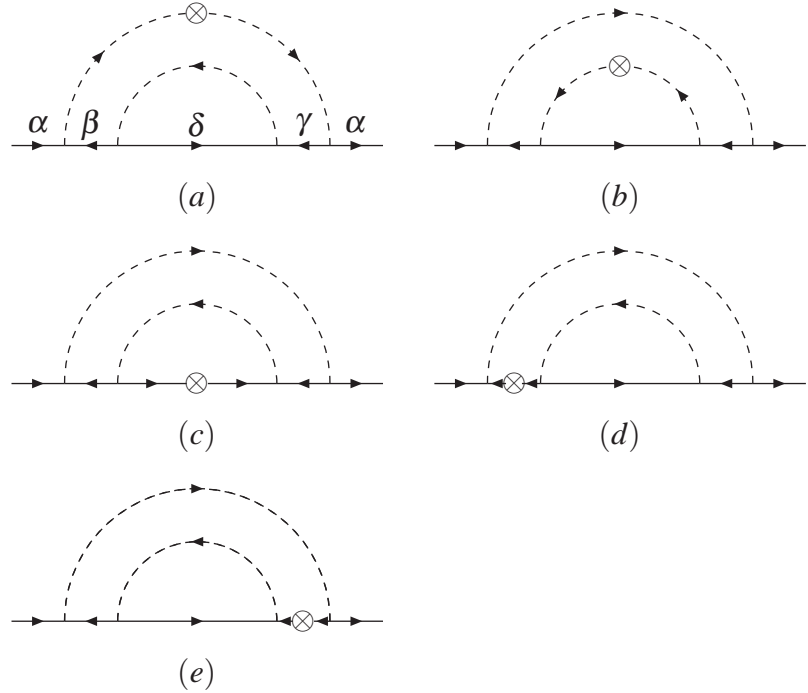
\Text(0,-40)[l]{Figure 1. Diagrams contributing to EDM of
$\ell_\alpha$.}

\end{picture}
\end{center}

It is interesting that the above form does not vanish even in the
case of two flavors where only a single Majorana phase can appear.
To see the point, it suffices to consider the easier case of type
II seesaw in eq. (\ref{eqn_seesaw}) with
\begin{eqnarray}
V=\left(\begin{array}{cc} c&s\\-s&c
\end{array}\right)\left(
\begin{array}{cc}u&\\&1\end{array}\right),
\end{eqnarray}
where $c$, $s$ are the cosine and sine of the mixing angle, and $u$
is the CP phase. Then, we find for instance
\begin{eqnarray}
(2|v_3|)^42i\Im[b^*_{\mu e}b_{ee}(b^\dagger b)_{e\mu}]
=c^2s^2(m_1^2-m_2^2)m_1m_2(u^2-u^{*2})
\end{eqnarray}
which does not vanish in general. This is a feature pertaining to
the Majorana-type couplings of charged leptons in eq.
(\ref{eqn_yukawa}) or the Majorana nature of neutrinos in type II
seesaw.

We will see in the next section that the combination
$b^\ast_{\delta\beta}b_{\delta\gamma}$ is no less constrained than
$(b^\dagger b)_{\beta\gamma}$. It is thus a good approximation to
keep the leading term at $r_\delta=0$ while ignoring small
corrections that are at most of order $r_\delta\ln r_\delta$. The
final answer for $d_\alpha$ thus looks like
\begin{eqnarray}
d_\alpha=C\frac{em_\alpha}{m^2}
\Im[b^\ast_{\gamma\alpha}b_{\beta\alpha}(b^\dagger b)_{\beta\gamma}]
[f(r_\beta,r_\gamma;0)-f(r_\gamma,r_\beta;0)]
\end{eqnarray}
where $C$ is a loop factor. This result entails an interesting sum
rule
\begin{eqnarray}
\frac{d_e}{m_e}+\frac{d_\mu}{m_\mu}+\frac{d_\tau}{m_\tau}=0 %
\label{eqn_sum}
\end{eqnarray}
which is exact up to small relative corrections of
$O(r_{\alpha,\delta})$. And up to logarithmic enhancements, we have
approximately,
\begin{eqnarray}
d_\alpha\sim C\frac{em_\alpha(r_\beta-r_\gamma)}{m^2}
\Im[b^\ast_{\gamma\alpha}b_{\beta\alpha}(b^\dagger b)_{\beta\gamma}]
\end{eqnarray}

We are now ready to present the results for the graphs. Graph (a) is
symmetric in $\beta$ and $\gamma$, and does not contribute to EDM.
Graphs (b) and (c) each contain symmetric and antisymmetric terms,
while the sum of (d) and (e) is antisymmetric. The contribution to
EDM is
\begin{eqnarray}
d_\alpha=\frac{2^5em_\alpha}{24(4\pi)^4m^2}
\Im\big[b^*_{\gamma\alpha}b_{\beta\alpha}(b^\dagger
b)_{\beta\gamma}\big]J(r_\beta,r_\gamma)%
\label{eqn_d}
\end{eqnarray}
where again summation over $\beta,~\gamma$ is implied and $J$ is a
sum over four graphs:
\begin{eqnarray}
J(r_\beta,r_\gamma)=J^{(b)}(r_\beta,r_\gamma)
+J^{(c)}(r_\beta,r_\gamma)+J^{(d+e)}(r_\beta,r_\gamma)
\end{eqnarray}
Each of these four graphs has ultraviolet sub-divergences. In
$4-2\epsilon$ dimensions, they are
\begin{eqnarray}
J^{(b){\rm div}}=-2J^{(c){\rm div}}=-2J^{(d+e){\rm div}}
=F(r_\beta,r_\gamma)\Gamma(\epsilon),
\end{eqnarray}
where the arguments in $J$ are suppressed and
\begin{eqnarray*}
F(b,c)&=&-\frac{2[b^2+c^2-bc-bc(b+c)+b^2c^2]}
{(b-1)^2(c-1)^2(b-c)}\\
&&+\frac{b^2[-3c+b(1+b+c)]\ln b}{(b-1)^3(b-c)^2}
-\frac{c^2[-3b+c(1+b+c)]\ln c}{(c-1)^3(b-c)^2}
\end{eqnarray*}
The divergences are canceled on summation as they must be.

The analytic result for the finite part is much more lengthy. In
addition to the displayed function $F$, each graph involves one or
two other twofold parameter integrals that can be worked out in
terms of the fractions, logarithms $\ln r_\beta$ and $\ln r_\gamma$,
and the dilogarithms ${\rm Li}_2(1-r_\beta)$ and ${\rm
Li}_2(1-r_\gamma)$. We will not record these exact results but the
sum of all graphs that has been expanded to the leading order in
$r_{\beta,\gamma}$:
\begin{eqnarray}
J(r_\beta,r_\gamma)=\frac{r_\beta^2+r_\gamma^2-r_\beta r_\gamma}
{r_\beta-r_\gamma} %
+\frac{r_\beta^2(r_\beta-3r_\gamma)\ln r_\beta
-r_\gamma^2(r_\gamma-3r_\beta)\ln r_\gamma} {2(r_\beta-r_\gamma)^2}
+\cdots,%
\label{eqn_J}
\end{eqnarray}
where the dots stand for higher order terms in $r_{\beta,\gamma}$.
Since the charged lepton masses are hierarchical, further expansion
is possible; for $1\gg r_\beta\gg r_\gamma$, we have
\begin{eqnarray}
J(r_\beta,r_\gamma)=
r_\beta-2r_\gamma+\frac{1}{2}(r_\beta-3r_\gamma)\ln r_\beta+\cdots
\end{eqnarray}
We have tested that the leading terms shown in eq. (\ref{eqn_J})
recover the first three digits of the exact results at $m=200~\GeV$
and are good enough for our later numerical analysis.

\section{Numerical analysis}
\label{sec:num}

Our result for the charged lepton EDMs shown in eqs.
(\ref{eqn_d},\ref{eqn_J}) is suppressed by charged lepton masses
squared over four powers of the scalar mass, and has a mild
logarithmic enhancement factor. This is a parametrically large
contribution. For instance, at our reference point $m=200~\GeV$, we
have $J(r_e,r_\mu)\approx 1.83\times 10^{-6}$, $J(r_e,r_\tau)\approx
J(r_\mu,r_\tau)\approx 2.94\times 10^{-4}$, and
\begin{eqnarray}
d_e\sim 4\times 10^{-30}\times[b~{\rm factors}]~e~\cm
\end{eqnarray}
which would be within the reach in the next generation of experiment
at the sensitivity of order $10^{-31}~e~\cm$ \cite{Kawall:2004nv}.

However, the same Yukawa couplings in eq. (\ref{eqn_yukawa})
induce other effects as well, and a realistic estimate of EDM
should take into account the constraints from those effects. In
this section, we present our numerical results in two approaches.
The main constraints considered are from lepton flavor violating
(LFV) decays of charged leptons.  Also mentioned are anomalous
magnetic moments and $0\nu 2\beta$ decays. We start with a model
independent analysis in the next subsection and then specialize to
the case of type II seesaw. The constraints in the second approach
are more stringent because of less free parameters involved.

\subsection{Approach 1: model independent result}

The Yukawa couplings in eq. (\ref{eqn_yukawa}) mediate radiative LFV
decays at one loop level and purely leptonic decays at tree level.
The branching ratio for the radiative decay is
\begin{eqnarray}
\Br(\ell_\beta\to\ell_\alpha\gamma)=\frac{3^3\alpha}{2^6\pi}%
\left|\frac{(b^\dagger b)_{\alpha\beta}}{G_Fm^2}\right|^2 %
B_\beta B_\xi,%
\label{eqn_br_rad}
\end{eqnarray}
with $B_\mu=1$ and $B_\tau\approx 17\%$. $B_\xi$ is a model
parameter which equals $(8/9)^2$ for the contribution of
$\xi^{\pm\pm}$ alone (in this subsection) and equals $1$ when both
$\xi^{\pm\pm}$ and $\xi^{\pm}$ are included as in type II seesaw
model (in the next). The branching ratio for the purely leptonic
decay is
\begin{eqnarray}
\Br(\ell_\delta\to\bar\ell_\alpha\ell_\beta\ell_\gamma)
=\frac{1}{2^2}\left|\frac{b_{\delta\alpha}b_{\beta\gamma}}{G_Fm^2}\right|^2
(2-\delta_{\beta\gamma})B_\delta,%
\label{eqn_br_lep}
\end{eqnarray}
which is only induced by $\xi^{\pm\pm}$ exchange. The factor
$(2-\delta_{\beta\gamma})$ distinguishes between identical and
nonidentical particles in the final state. Using the experimental
bounds on the branching ratios we can constrain $|(b^\dagger
b)_{\alpha\beta}|/(G_Fm^2)$ and
$|b_{\delta\alpha}b_{\beta\gamma}|/(G_Fm^2)$ respectively. These
numbers are shown in table 1, and will be employed to set
conservative upper bounds on EDMs.

\begin{table}
\begin{center}
\begin{tabular}{|c|l|l|l|l|l|}
\hline modes & $\mu\to e\gamma$ & $\tau\to e\gamma$ %
& $\tau\to\mu\gamma$ & $\mu\to 3e$ & $\tau\to 3e$\\
\hline Br  &$1.2~10^{-11}$ \cite{Brooks:1999pu} %
& $1.1~10^{-7}$ \cite{Aubert:2005wa} %
& $4.5~10^{-8}$ \cite{Hayasaka:2007vc} %
& $1.0~10^{-12}$ \cite{Bellgardt:1987du} %
& $4.3~10^{-8}$ \cite{Aubert:2007pw}\\
\hline %
bounds & $1.2~10^{-4}$ & $2.9~10^{-2}$ & $1.9~10^{-2}$ &
$2.0~10^{-6}$ & $1.0~10^{-3}$ \\
\hline %
\hline modes & $\tau\to 3\mu$ & $\tau\to\bar e 2\mu$ %
& $\tau\to\bar\mu 2e$ & $\tau\to\bar ee\mu$ & $\tau\to\bar\mu\mu e$\\
\hline Br  &$5.3~10^{-8}$ \cite{Aubert:2007pw} %
& $5.6~10^{-8}$ \cite{Aubert:2007pw} %
& $5.8~10^{-8}$ \cite{Aubert:2007pw} %
& $8.0~10^{-8}$ \cite{Aubert:2007pw} %
& $3.7~10^{-8}$ \cite{Aubert:2007pw}\\
\hline %
bounds & $1.1~10^{-3}$ & $1.1~10^{-3}$ & $1.2~10^{-3}$ &
$9.7~10^{-4}$ & $6.6~10^{-4}$ \\
\hline %
\end{tabular}
\caption{Experimental upper bounds on branching ratios of decays in
eqs. (\ref{eqn_br_rad}, \ref{eqn_br_lep}) set upper bounds on
$|(b^\dagger b)_{\alpha\beta}|/(G_Fm^2)$ and
$|b_{\delta\alpha}b_{\beta\gamma}|/(G_Fm^2)$ respectively.}%
\label{tab_1}
\end{center}
\end{table}

Each $d_\alpha$ has three terms proportional to $J(r_e,r_\mu)$,
$J(r_e,r_\tau)$, and $J(r_\mu,r_\tau)$ respectively, for instance,
\begin{eqnarray}
\frac{96\pi^4}{G_F^2}\frac{d_e}{em_e}&=&%
+\Im\left[\frac{b^*_{\mu e}b_{ee}}{m^2G_F}%
\frac{(b^\dagger b)_{e\mu}}{m^2G_F}\right]m^2J(r_e,r_\mu)
\nonumber\\
&&+\Im\left[\frac{b^*_{\tau e}b_{ee}}{m^2G_F} %
\frac{(b^\dagger b)_{e\tau}}{m^2G_F}\right]m^2J(r_e,r_\tau)
\nonumber\\%
&&+\Im\left[\frac{b^*_{\tau e}b_{\mu e}}{m^2G_F} %
\frac{(b^\dagger b)_{\mu\tau}}{m^2G_F}\right]m^2J(r_\mu,r_\tau)
\end{eqnarray}
The first term is much smaller because of a smaller $J$ factor and
more severely suppressed moduli of the products of $b$ factors, and
can safely be ignored. In the optimistic case where the products of
$b$ factors in the last two terms are purely imaginary and add
constructively, we get at $m=200~\GeV$,
\begin{eqnarray}
|d_e|\le 8.1\times 10^{-35}~e~\cm
\end{eqnarray}
Since our bounds in table 1 are given independently of $m^2$ while
$m^2J$ depends only logarithmically on $m^2$, the above bound is
stable against mild variations of $m$. Similarly, we obtain
\begin{eqnarray}
|d_\mu|\le 1.4\times 10^{-32}~e~\cm
\end{eqnarray}
The expression for $d_\tau$ contains several combinations of $b$
factors that cannot be constrained in LFV decays, so that a direct
bound is not possible. But we can utilize the sum rule
(\ref{eqn_sum}) to set a bound
\begin{eqnarray}
|d_\tau|\le 5.2\times 10^{-31}~e~\cm
\end{eqnarray}
The limit on $|d_e|$, though larger than the four-loop SM result
\cite{Pospelov:1991zt} and the bounds reached via other mechanisms
\cite{Ng:1995cs,Archambault:2004td,Chang:2004pba,deGouvea:2005jj},
is still about three orders of magnitude below the precision
reachable in the near future \cite{Kawall:2004nv}.

\subsection{Approach 2: a case study in type II seesaw}

The discussion in the previous subsection is model independent. When
the Yukawa interaction in eq. (\ref{eqn_yukawa}) is part of a
complete structure in a model, more stringent constraints on EDMs
can be obtained. This is the case particularly in the type II seesaw
model where the Yukawa couplings are related via eq.
(\ref{eqn_seesaw}) to the neutrino masses and mixing matrix which
have been determined to certain extent. In this subsection we will
not attempt a global fitting but demonstrate the point by a case
study in this model.

The mixing pattern determined by oscillation data is close to the
tribimaximal texture \cite{Harrison:2002er}. We will work in this
simplified scenario. There is then no Dirac phase but there can be
two Majorana phases $u_{1,2}$:
\begin{eqnarray}
V=\left(\begin{array}{ccc}
\sqrt{\frac{2}{3}}u_1&\frac{1}{\sqrt{3}}u_2&0\\
-\frac{1}{\sqrt{6}}u_1&\frac{1}{\sqrt{3}}u_2&\frac{1}{\sqrt{2}}\\
\frac{1}{\sqrt{6}}u_1&-\frac{1}{\sqrt{3}}u_2&\frac{1}{\sqrt{2}}
\end{array}\right)
\end{eqnarray}
Then, the matrix $b^\dagger b$ is real and symmetric,
\begin{eqnarray}
4|v_3|^2b^\dagger b
&=&\frac{1}{3}(m_1^2+m_2^2+m_3^2)1_3\nonumber\\
&&+\frac{1}{6}\left(\begin{array}{ccc}
2\Delta_{13}&-2\Delta_{12}&2\Delta_{12}\\
-2\Delta_{12}&-\Delta_{13}&-\Delta_{13}-2\Delta_{23}\\
2\Delta_{12}&-\Delta_{13}-2\Delta_{23}&-\Delta_{13}
\end{array}\right)
\end{eqnarray}
with $\Delta_{ij}=m^2_i-m^2_j$.

The off-diagonal moduli $|(b^\dagger b)_{\alpha\beta}|$ depend
explicitly on $\Delta_{ij}$, which have been determined e.g. in
\cite{Garayoa:2007fw} to be, $\Delta_{21}=7.6\times 10^{-5}~\eV^2$,
$|\Delta_{31}|=2.4\times 10^{-3}~\eV^2$. The bound on $\Br(\mu\to
e\gamma)$ then implies (using $B_\xi=1$)
\begin{eqnarray}
|v_3|^2m^2G_F>5.75\times 10^{-2}~\eV^2%
\label{eqn_bound1}
\end{eqnarray}
Since $\Br(\tau\to e\gamma)$ also depends on $\Delta_{12}$, its less
stringent bound is useless. Instead, its relation to $\Br(\mu\to
e\gamma)$ in type II seesaw and the experimental bound on the latter
mean
\begin{eqnarray}
\Br(\tau\to e\gamma)=B_\tau\Br(\mu\to e\gamma)\le 2.0\times
10^{-12}
\end{eqnarray}
which is much below the current bound. We also notice that the bound
on $\Br(\tau\to\mu\gamma)$, though more than three orders of
magnitude larger than $\Br(\mu\to e\gamma)$, gives a constraint that
is only slightly weaker than in eq. (\ref{eqn_bound1}), because of
an enhancement factor $|\Delta_{31}|/\Delta_{21}$. A similar
relation also holds between $\Br(\tau\to 3e)$ and $\Br(\mu\to 3e)$;
the more stringent bound on the latter implies
\begin{eqnarray}
\Br(\tau\to 3e)=B_\tau\Br(\mu\to 3e)\le 1.7\times 10^{-13}
\end{eqnarray}
which is much smaller than its current bound and thus more
difficult to observe than other decay modes of $\tau$.

To proceed further with leptonic decays of $\tau$ and EDM, we set
all neutrino masses in $b^*_{\delta\alpha}b_{\beta\gamma}$ (but not
in $(b^\dagger b)_{\beta\alpha}$ of course) to be equal to their
average value $\bar m_\nu$. This simplification holds true barring
very delicate cancellation among neutrino mass differences and
Majorana phases. Then, the most stringent constraints on $\mu\to
3e$, $\tau\to\bar ee\mu,~\bar\mu\mu e$ (together with the less
stringent one on $\tau\to\bar e2\mu$) are proportional to
$|u_1^2-u_2^2|$. We may reach the most optimistic values of EDMs by
assuming $u_1^2=u_2^2=e^{i\phi}$ to avoid these bounds. The
remaining ones on $\tau\to 3\mu,~\bar\mu 2e$ yield comparable
constraints:
\begin{eqnarray}
\frac{\bar m_\nu^2|\sin\phi|}{8|v_3|^2m^2G_F} <1.1\times 10^{-3}, %
~\frac{\bar m_\nu^2|\sin(\phi/2)|}{4|v_3|^2m^2G_F}<1.2\times 10^{-3}%
\label{eqn_bound2}
\end{eqnarray}
The electron EDM being proportional to $\Im(u_1^{*2}u_2^2)$
vanishes, while the other two simplify to
\begin{eqnarray}
\frac{d_\mu}{em_\mu}=-\frac{d_\tau}{em_\tau} %
=\frac{\bar m_\nu^2\Delta_{13}\sin\phi}{2^{11}~3\pi^4|v_3|^4m^2}
J(r_\mu,r_\tau),
\end{eqnarray}
barring cancellation of $O(\Delta_{21}/|\Delta_{31}|)\sim 3\%$ or
$O([\Delta_{21}J(r_e,r_\mu)]/[|\Delta_{31}|J(r_\mu,r_\tau)])\sim
2\times 10^{-4}$. The bounds in eqs. (\ref{eqn_bound1},
\ref{eqn_bound2}) then give at $m=200~\GeV$
\begin{eqnarray}
|d_\mu|<2.0\times 10^{-33}~e~\cm,~|d_\tau|<3.4\times
10^{-32}~e~\cm
\end{eqnarray}
As expected, this result is better than the model-independent one
in the previous subsection.

In this special scenario, the effective neutrino mass measured in
$0\nu 2\beta$ decays is $m_{\beta\beta}=(2m_1+m_2)/3\sim \bar
m_\nu$. The contributions to anomalous magnetic dipole moments
depend on the diagonal elements of $b^\dagger b$ and are given by
\begin{eqnarray}
a_e&=&\frac{3}{(4\pi)^2}\frac{\bar
m_\nu^2m_e^2G_F}{4|v_3|^2m^2G_F}<
1.1\times 10^{-14}\nonumber\\
a_\mu&=&\frac{3}{(4\pi)^2}\frac{\bar
m_\nu^2m_\mu^2G_F}{4|v_3|^2m^2G_F}<4.7\times 10^{-10}
\end{eqnarray}
using eq.(\ref{eqn_bound1}) and $\bar m_\nu\sim 0.21~\eV$ from a
recent update of cosmological bounds on the sum of neutrino masses
\cite{Hannestad:2008js}. Both $a_e$ and $a_\mu$ are below the
potential gap between measurements and SM expectations
\cite{Odom:2006zz, Yao:2006px}.

\section{Conclusion}

CP violation in the lepton sector has remained an experimentally
unexplored issue. The charged lepton EDMs offer a potential arena to
detect it. This is especially encouraged by the experimental
precision in EDM measurements that has been reached and will
possibly be accessible. However, it has been found previously that
it is hard to obtain a not too tiny EDM for charged leptons from a
flavor CP source. This may be blamed on the very light, thus almost
degenerate at the electroweak scale, neutrinos. Together with a
small mixing angle out of three, this makes a Dirac phase
effectively unobservable; and it suppresses the effects of Majorana
phases on EDMs by several factors of neutrino masses. We have thus
been motivated to consider a CP source that arises from
Majorana-type Yukawa couplings of charged leptons. Such couplings
may appear naturally in SM with an extended scalar sector, such as
type II seesaw model, but we have presented our analytic results in
a general setting. We found that the EDMs so obtained are
parametrically large. They are only suppressed by charged lepton
masses squared over four powers of heavy scalar masses for order one
Yukawa couplings that may be naturally arranged, for instance, in
type II seesaw model by assigning a tiny vacuum expectation value
for the scalar triplet.

Nevertheless, the fate with a flavor CP source seems insurmountable.
While a large enough EDM, though flavor diagonal, demands reasonably
large flavor changing couplings, this may not be allowed by strictly
bounded LFV transitions. With these bounds taken into account, we
found that the electron EDM is at least three orders of magnitude
below the precision achievable in the near future, although it is
still much larger than the contributions considered previously.

\vspace{0.5cm}
\noindent %
{\bf Acknowledgement} This work is supported in part by the grants
NCET-06-0211 and NSFC-10775074.

\vspace{0.5cm}
\noindent %

\end{document}